# Energy-efficient picosecond spin-orbit torque magnetization switching in ferro and ferrimagnetic films

Eva Díaz[1], Alberto Anadón[1], Pablo Olleros-Rodríguez[2]*, Harjinder Singh[1], Héloïse Damas[1], Paolo Perna[2], Martina Morassi[3], Aristide Lemaître[3], Michel Hehn[1] and Jon Gorchon[1]*

[1] Université de Lorraine, CNRS, IJL, F-54000 Nancy, France
[2] IMDEA Nanociencia, 28049 Madrid, Spain
[3] Université Paris-Saclay, CNRS, Centre de Nanosciences et de Nanotechnologies, 91120 Palaiseau, France
* Corresponding authors: jon.gorchon@univ-lorraine.fr, pablo.olleros@imdea.org

Electrical current pulses can be used to manipulate magnetization efficiently via spin-orbit torques (SOTs). Pulse durations as short as a few picoseconds have been used to switch the magnetization of ferromagnetic films, reaching the THz regime. However, little is known about the reversal mechanisms and energy requirements in the ultrafast switching regime. In this work, we quantify the energy cost for magnetization reversal over 7 orders of magnitude in pulse duration, in both ferromagnetic and ferrimagnetic samples, bridging quasi-static spintronics and femtomagnetism. To this end, we develop a method to stretch picosecond pulses generated by a photoconductive switch by an order of magnitude. Thereby, we can create current pulses from picoseconds to durations approaching pulse width available with commercial instruments. We show that the energy cost for SOT switching decreases by more than an order of magnitude in all samples when the pulse duration enters the picosecond range. We project an energy cost of 9 fJ for a 100 x 100 nm² ferrimagnetic device. Micromagnetic and macrospin simulations unveil a transition from a non-coherent to a coherent magnetization reversal with a strong modification of the magnetization dynamical trajectories as pulse duration is reduced. Our results cement the potential for high-speed magnetic spin-orbit torque memories and highlights alternative magnetization reversal pathways at fast time scales.

Achieving both *fast* and *energy-efficient* control of magnetization via electrical means has long been a major goal in the field of spintronics[1–3]. The spin-orbit torque (SOT) effect exploits the spin-orbit coupling in a non-magnetic layer to convert charge currents into spin currents, and then torque and switch the magnetization of an adjacent magnetic layer[4–6]. Now, SOT is a leading candidate for the next generation of magnetic memories due to the associated speed and device durability[7].

Fast SOT switching was demonstrated first in a prototypical 200-nm wide out-of-plane magnetized memory bit of Co with electrical pulses going down to 200 picoseconds (ps) in duration[8]. The critical current ($I_c$) for switching shows two different trends depending on pulse duration ($\tau_p$)[9–11]: in the long pulse regime, $I_c$ exhibits a slow $-\log(\tau_p)^{0.5}$ law and switching is attributed to a thermally-activated mechanism[12]. On the other hand, in the short pulse regime, $I_c$ exhibits a divergent $1/\tau_p$ law normally explained by a quasi-deterministic reversal by domain nucleation and domain wall (DW) propagation requiring a constant amount of angular momentum[8,11,13]. In this fast regime the role of heating is generally assumed to be much more limited[8,11,14], despite some recent works showing its importance in the switching dynamics[10]. Such a $1/\tau_p$ tendency, extrapolated to pulses of a few ps in duration, hinted to current densities on the order of $10^{14}$ A m$^{-2}$. This would lead to substantial heating and energy consumption, making such ultra-fast regime impractical.

However, recently, SOT switching of a Co film was demonstrated using a single 6 ps electrical pulse[15]. Sub-threshold time-resolved dynamics confirmed the presence of SOT acting at the ps time scale and hinted at the presence of non-negligible thermal effects[15]. Notably, the devices were large 4 x 5 µm² films, which casted doubts on the plausibility of a DW propagation mechanism. Indeed, huge DW velocities of ~$10^4$ m/s would be necessary for DWs to cover the width of the sample in 6ps[16,17]. Furthermore, switching revealed a zero-crossing of the out-of-plane magnetization component ~70 ps after the electrical pulse arrival[18]. A post-pulse zero-crossing is inconsistent with the classical DW nucleation and propagation picture, as a driving force is required. A post-pulse reversal would be instead in good agreement with predictions obtained from macrospin models for ps pulses[15,18]. Indeed, as previous works show[15,18–20], short and intense excitations can bring the magnetization away from its equilibrium position, leading to precession around the effective field. Depending on damping and effective field, the magnetization can potentially reach the equator and relax into the reversed direction, or even over-oscillate back into the original direction[20]. Importantly, in this picture, magnetization dynamics are fully coherent, in sharp contrast with observations made to this day with pulses longer than 200 ps[10,13,21,22]. This interpretation would therefore require a crossover from a non-coherent to a coherent reversal between the ns and ps time scales. This change of regime should be made evident by a departure of $I_c$ from the $1/\tau_p$ law, but so far, critical current densities and energy costs for SOT in the ps regime have never been characterized, and only rough estimations exist[15].

In this work, we report SOT switching for a wide range of pulse durations, from the μs scale down to the ps scale. Three different samples were investigated, either ferromagnetic or ferrimagnetic. By characterizing current densities and energies we demonstrate that the critical current density departs from the $1/\tau_p$ law when entering the ps regime and that switching energy decreases by least an order of magnitude for all three samples. Macrospin and micromagnetic simulations reveal a transition from a non-coherent to a coherent reversal into the ps regime, and confirm both the higher energy efficiency at short time scales and the important role of temperature.

**Devices and ps-pulse generation/detection platform**

The following stacks were studied: Ta(4)/Pt(4)/Co(1)/Cu(1)/Ta(4)/Pt(1), Ta(10)/CoFeB(1)/MgO(1)/Pt(2), and Ta(3)/Pt(5)/Co(1.4)/Gd(1)/Ta(2)/Pt(2), nicknamed *Co*, *CoFeB*, and *CoGd*, respectively. Numbers represent thicknesses in nm. All samples present perpendicular magnetic anisotropy. *Co* and *CoFeB* are conventional ferromagnets, extensively studied in transport experiments, while *CoGd* is a ferrimagnet, in which both SOT switching[22] and all-optical magnetization switching[23] have been reported.

The devices shown in Fig. 1a, 1c and insets were fabricated by a multi-step lithography process onto LT-GaAs substrates (Methods). Devices include generator and detector photoconductive switches[24,25] excited by femtosecond laser beams (Methods) with the former biased with voltage $V_b$. Proper pulse transmission was ensured with coplanar waveguides (CPW) and impedance matched magnetic load terminations. We were able to generate 7-ps pulses up to tens of V in amplitude (Fig. 1b), which were calibrated using on-chip THz detectors (see Extended Data Fig. 1) using a previously developed method[25]. In order to stretch the pulses further, either the optical pulse has to be stretched to the desired pulse length (> 7 ps), which is not possible in our setup, or the recombination time of the substrate material has to be changed. The recombination time can be modified by changing the material altogether, but this requires the growth and characterization of new substrates and the fabrication of new devices. In order to extend the recombination time on the existing devices, we decided to exploit a vulnerability of most photoswitches: the lowering of the bandwidth with increasing fluence[26]. By increasing the pump power, we observe an increase in the decay time of our electrical pulses (Fig. 1b), allowing us access to pulses from 7 ps to 60 ps in FWHM, albeit with asymmetric profiles (Extended Data Fig. 2). Pulses between 0.5 ns and 1000 ns in duration were generated (Extended Data Fig. 3) using a commercial pulse generator and injected into the sample by bypassing the photoconductive switch with 40-GHz bandwidth GSG probes, as shown in Fig. 1c. The comparison of both typical photoswitch-generated and a 5-ns pulse is shown in Fig. 1b and 1d, respectively, detected on-chip via our THz detector in absolute units. Finally, we monitored the sample magnetic state using a home-built magneto-optical Kerr effect (MOKE) microscope (Methods).

**Threshold currents and in-plane field**

For all samples we increased the amplitude of the pulses until we ensured that a single pulse could induce deterministic SOT switching (Methods). Various after-pulse MOKE images were then recorded for different current densities around the threshold value. In this way, we measured the averaged switched area as a function of current (Suppl. Mat. S3), as shown in Fig. 2a for the case of the *Co* sample. The switching threshold $I_c$ was defined as the current $I$ (Methods) for which a switched area of 100 % was observed.

As shown in Fig. 2a, the 0-to-1 transition of the switched area is less pronounced for ns switching than for the ps one. Notably, around the threshold, the resulting state differs greatly from trial to trial (Extended Data Fig. 5), sometimes with multiple nucleation points, yielding a large standard deviation of the switched area. This observation confirms that stochasticity is at play in the ns regime, as has been suggested[10]. For ps switching, on the other hand, around the threshold a single switched domain consistently appears on the current injection side. These qualitative differences between the two regimes suggests that magnetization reversal is carried out through different mechanisms.

The same procedure was repeated for different in-plane fields in the *Co* sample (Fig. 2b, *CoFeB* and *CoGd* in Extended Data Fig. 6). When the in-plane field decreases, the initial magnetization tilt becomes smaller, requiring larger currents for the magnetization to reach the equator and switch. Such a behavior was observed in all regimes (ns, ps) and samples. Nevertheless, the impact of in-plane field intensity remains weak. At zero field, systematic switching was not observed, as expected, as an in-plane field is necessary to break the symmetry and enable SOT switching[27]. However, both ns and ps switching were achieved even at low fields (~20 mT), attainable by engineered dipolar fields or tilted anisotropy axis[28,29].

**Threshold current and energy as a function of pulse duration**

It is well accepted that for pulse durations well above the ns, magnetization reversal is thermally activated[12]. The magnetization is first tilted away from the out-of-plane direction due to SOT and, eventually, thermal excitations initiate reversal on defect sites and/or edges, which then expand via DW propagation. However, as the pulse approaches the ~ns mark, previous works have shown that domains consistently nucleate on the edges and then expand by deterministic edge-to-edge DW propagation[13,21,22]. For fast DWs, in the *flow* regime, DW velocity is proportional to the driving force[30]. It follows that for complete reversal, the DW has to propagate over the full sample length. As $\tau_p$ gets shorter, the critical current for complete reversal has therefore to increase as $I_c \propto 1/\tau_p$. In other words, the angular momentum necessary to reverse the magnetization remains constant. Consequently, the critical energy $E_c \propto I_c^2 \tau_p$ also follows a $1/\tau_p$ dependence. With this simple picture, a minimum of energy consumption is expected, assumed to be around the ~ns mark[8,11], which was confirmed in a CoFeB/MgO/CoFeB magnetic tunnel junction[9]. In contrast, more recent results have shown an ever-decreasing energy consumption as the pulse reaches the 200 ps mark[10]. Results with ferrimagnets suggest extreme energy-efficiency in nanoscale devices with pulses down to tens of ps[22]. Some works have suggested, alternatively, that ns pulse-switching may take place via multiple nucleation sites instead, for micron-sized devices[14].

Fig. 3a shows the evolution of the critical current $I_c$ as a function of $\tau_p$ from the µs scale to the ps scale. For reference, a rough estimation of the current density flowing the Pt layer within the *Co* stack $j_c^{Co}$ is indicated on the right axis (for details about calculation see Suppl. Mat. S7). Critical current increases slowly up until the ns regime, following a $-\log(\tau_p)^{0.5}$ law, as expected and observed in previous works[8,12]. As $\tau_p$ gets even shorter, we observe a good agreement with the $1/\tau_p$ law up to the hundreds of ps. For shorter ps pulses, however, we observe a significant deviation from the expected behavior, with a threshold current up to two orders of magnitude lower than a $1/\tau_p$ law extrapolation, for all studied samples. Fig. 3b shows the energy cost for switching $E$ (Methods), as a function of $\tau_p$. The evolution of the energy cost with $\tau_p$ also differs from the expected $1/\tau_p$ behavior. Instead of a sharp increase at short pulse durations, the energy consumption in the ps regime is actually *one order of magnitude lower than in the ns regime*. The *CoFeB* film requires slightly more energy at all $\tau_p$, despite its much lower damping parameter (Suppl. Mat. S9), certainly due to its much lower spin Hall angle (Suppl. Mat. S8). For the ferrimagnetic *CoGd*, current and energy thresholds are lower than for ferromagnetic samples, in line with previous observations[11,22]. Moreover, energy gains over the ferromagnetic counterparts seem to be slightly accentuated as $\tau_p$ is reduced. The energy efficiency of *CoGd* might be due to its low saturation magnetization, which reduces the amount of angular momentum necessary for switching[11]. On the right-hand axis of Fig. 3b we indicate the projected energy consumption for a 100 x 100 nm$^2$ device assuming a linear scaling[31], reaching a value as low as 9 fJ for *CoGd*.

Finally, all-optical switching (AOS) of the *CoGd* sample[32] was performed by shining the laser directly on the magnet. As indicated in Fig. 3b with a star, reversal was achieved using a single ~150 fs optical pulse of 2.5 mJ cm$^{-2}$, resulting in an absorbed energy of 55 pJ over a 3 x 4 µm$^2$ area (Suppl. Mat. S5). The switching mechanism in this case is purely due to the rapid electronic heating of the system, which triggers an angular momentum exchange between the Co and Gd layers resulting in a sub-ps reversal of the net magnetization[23,33]. We believe that this mechanism is probably also present when heating up the magnet with ps electrical pulses[34], which combined with SOT might help the magnetization reversal, resulting in an even higher energy efficiency.

**Modelling in the ps and ns regimes**

Previous works modelled SOT switching due to ps electrical pulses using a macrospin model coupled to a heat diffusion equation and temperature dependent parameters[15,18]. These simulations fitted the ultrafast dynamics reasonably well, however, magnetization reversal above the ns regime is known to be determined by domain nucleation and propagation[11] where a macrospin approach should fail. Therefore, in order to address the full pulse duration range from ns to ps, we developed a micromagnetic Mumax$^3$ code to perform simulations, which features a temperature-dependent anisotropy and stochastic thermal fields coupled to a thermal diffusion model[35]. The model predicts the temperature rise due to Joule heating, which is used to tune magnetic anisotropies and thermal fields (Suppl. Mat. S10). For comparison, temperature-dependent macrospin simulations were performed, including the temperature-dependences from Jhuria et al.[15], but with a fixed norm of M. In this numerical work, we focused our attention on the Co film, which we had best characterized (Suppl. Mat. S6 and S8).

As shown by simulations in Fig. 4a, SOT reversal induced by 2 ns pulses is governed by multiple nucleation events all over the sample area, which then expand due to SOT [14]. This reversal mechanism is in stark contrast to the edge-to-edge DW propagation reversal observed in other works[10,13,21]. Further analysis revealed two important factors: device dimensions and magnetic edges. In our samples, current is injected in the middle of the magnetic strip (Extended Data Fig. 7, 8) and, therefore, no SOT is exerted on the tilted edge-magnetization which favors a single edge-domain nucleation[13]. Moreover, the large dimensions make an edge-to-edge propagation challenging, as current would need to increase (to speed up DWs), leading to more nucleation (Extended Data Fig. 9). For these reasons, in our simulations, reversal is dominated by nucleation, with only a small contribution from domain growth.

We then performed simulations with shorter, 6-ps wide, electrical pulses, with identical parameters. As depicted in Fig. 4b, the reversal becomes almost uniform and coherent. The degree of coherence is highlighted in the time evolution of the norm of the magnetization (right axis on Fig. 4c-d), defined as $|\boldsymbol{m}| = \sqrt{m_x^2 + m_y^2 + m_z^2}$, where $m_{x,y,z}$ is the spatially-averaged (x,y,z)-component of the reduced magnetization. A value of $|\boldsymbol{m}| \sim 1$ is observed in the ps switching case all throughout the switching process (Fig. 4d), implying a parallel alignment of all the spins in the film. On the other hand, an important drop in $|\boldsymbol{m}|$ is observed for the ns switching case (Fig. 4c), indicating a strong misalignment of spins within the film, and so, the presence of magnetic domains.

Another point to highlight is the reversal timing predicted by both macrospin and micromagnetic simulations. In agreement with previous reports[14], the reversal induced by the ns pulse takes place entirely during current pulse injection (Fig. 4c). In contrast, the reversal induced by the ps pulse happens well after the pulse (Fig. 4d), in agreement with recent results[18]. The ns macrospin reversal is earlier than the ns micromagnetic one (Fig. 4c) because of higher threshold currents in the macrospin case, as it does not benefit from the thermal fields.

Despite using the same parameters in both pulse duration regimes, the paths taken for reversal are outstandingly different. In Fig. 4e-h, we plot in yellow, rose and dark blue colors the dynamics from 0 to the pulse peak (arrival) time $t_0$, from $t_0$ to the magnetization switching time $t_{SW}$ (time for which magnetization crosses zero), and for $t > t_{sw}$, respectively. As shown in Fig. 4e for the 2-ns pulse, the macrospin dynamics predict a magnetization that is initially tilted towards the external field direction ($+x$), until SOT pulls it towards the spin density $\sigma_y$, reaching the equator. In the micromagnetic simulations (Fig. 4f), a sharp change can be observed (in rose color), which corresponds to the moment in which energy barriers are low enough for nucleation to take place. This results in a sudden decrease of the film total magnetization (as shown by $|m|$ in Fig. 4c). On the contrary, for short and intense pulses (Fig. 4g), the magnetization is initially torqued away from its equilibrium position, towards the spin density $\sigma_y$, resulting in initial dynamics almost *orthogonal* to the ns case. The micromagnetic simulation (Fig. 4h) show very similar dynamics. After the current is turned off, in both ps macrospin and micromagnetic simulations (Fig. 4g and 4h, respectively), the magnetization swings around the effective field until it reaches the equator (in rose color). Once there ($t = t_{sw}$), the effective field changes drastically due to the anisotropy field flipping, and the magnetization precesses towards its new reversed equilibrium (dark blue color on Fig. 4e-h). In the ps case, the anisotropy field flipping changes the apparent precession direction from a top-view perspective (change from rose to dark blue in Fig. 4g-h). In the micromagnetic simulation, for the ns case (Fig. 4f), domain formation prevents such precession from taking place.

Independent of the parameters (Methods), we find that in all micromagnetic simulations a smaller energy consumption is required for the ps regime. In Fig. 5 we plot the estimated threshold current densities and energies as a function of $\tau_p$, simulated with the same set of parameters as in Fig. 4. Additionally, we plot macrospin estimations both with and without temperature dependences. In all instances, we observe a large gain energy-wise, and with temperature effects included, we find reasonably close values for the thresholds. However, the thermal simulations are described by a $1/\tau_p^{0.5}$ law, which does not fit the experimental data. Particularly, there is an important discordance in the 100 – 500 ps region, which happens to be close to the ferromagnetic resonance (FMR) period of the system. The assumed thermal dependences are most likely not right and lead to the observed discrepancy. The issue may also be related to the very different temporal shapes of the experimental pulses (Methods). In Fig. 5 we also highlight the crossing point of the $1/\tau_p$ law and the *athermal* macrospin model (marked by a yellow circle). This point marks the upper limit of the $1/\tau_p$ regime, since if the macrospin threshold is surpassed a coherent reversal will necessarily be favored. Accounting for temperature effects, the coherent limit should be favored at lower currents.

Lastly, we discuss the possible reasons behind the energy efficiency in the ps regime. In order to reverse the magnetization, it is necessary to supply at least enough angular momentum to bring the magnetization onto the hard-plane (top of the energy barrier). Such an amount is expected to be fixed, indeed suggesting a $1/\tau_p$ dependency should exist. This expectation seems to fit in the ns regime, but breaks down as the pulse gets shorter. Two arguments might explain this: first, a fast heating of the sample could significantly modify the energy landscape, reducing barriers and opening new energy-efficient pathways for magnetization reversal[15,18]. Second, it is likely that in the ns regime angular momentum is wasted either in nucleation or in the domain wall propagation process, such as during the initial wall transformation[11]. For short enough pulses, a coherent, almost-adiabatic, reversal is likely more energy efficient than a lossy domain nucleation and propagation scheme. We postulate that the characteristic time scale of the transition from one regime to another must be related to the FMR period and/or the thermal diffusion time constant. Alternatively, the transition time scale could be related to the average time for a domain wall to reach a neighboring nucleation center, and might therefore depend on defect densities, device dimensions, nucleation probabilities and/or domain wall speeds. A sample size dependent study might show different energy trends.

To finish, we speculate on the possibility of speeding up the reversal from tens of ps[18] to a few ps. With enough current (above threshold), SOT could bring the magnetization to the equator within the pulse duration, resulting in an ultrafast zero crossing. The magnetization would then relax to the opposite magnetic state governed by anisotropy fields and damping. However, such high current density may: a) no longer be energy efficient and, b) lead to damage. An alternative way to speed up the reversal could be to exploit field-like torques. Indeed, for an equal damping, field-like torques may drive the magnetization directly to the extreme of the opposite hemisphere, faster than with damping-like torques[36].

**Conclusions**

We have demonstrated the energy efficiency of SOT-induced magnetization reversal in the ps regime, in both ferromagnetic and ferrimagnetic samples, by systematically characterizing critical currents and energy costs. We report a projected energy consumption of 9 fJ for electrical switching of 100-nm devices, 10x lower than for both ns-SOT and AOS. These results unveil a deviation from previously accepted models based on nucleation and propagation mechanisms. Micromagnetic modelling confirms the existence of a crossover from a non-coherent to a coherent switching. The model also correctly predicts the decrease in energy consumption as pulse duration gets shorter. We believe that coherent switching is made energetically favorable over nucleation and propagation in ps time scale due to the ultrafast injection of spin—and heat—that minimizes losses to the lattice and substrate, while efficiently driving the magnetization. Now, size dependent studies, down to application-level dimensions around ~100 nm are needed to check if the same energy savings are present or not in the ps regime. These results show that SOT remains an excellent candidate for further developing both *fast* and *energy-efficient* magnetic memories and logic devices. Our methods could be used to study spin-torques in antiferromagnetic systems which have naturally fast dynamics.

**Acknowledgments**
We wish to thank particularly Laurent Badie and Stéphane Suire, and more generally the CC-MINALOR, CC-HERÉ and CC-MAGNETISM platforms for their technical support on cleanroom aspects and for the fabrication of mechanical components for the setup. We thank Sébastien Petit-Watelot for useful discussions. This work was supported by the ANR Project Nos. ANR-20-CE24-0003 SPOTZ and ANR-20CE09-0013 UFO, by the impact project LUE-N4S part of the French PIA project "Lorraine Université d'Excellence," Ref. No. ANR-15IDEX-04-LUE, by the Région Grand Est and the Metropole Grand Nancy, for the Chaire PLUS and the "FEDERFSE Lorraine et Massif Vosges 2014-2020," a European Union Program, and by the French RENATECH network. This work was also supported by the France 2030 government investment plan managed by the French National Research Agency under grant references PEPR SPIN – TOAST ANR-22-EXSP-0003 and SPINMAT ANR-22-EXSP-0007. P.O.-R. and P.P. acknowledge support from Spanish AEI/MICINN through projects PID2021-122980OB-C52 (ECLIPSE-ECoSOx), CNS2022-136143 (SPINCODE) and CEX2020-001039-S. We acknowledge NVIDIA Corporation for the GPU Quadro P6000 donated for carrying out the micromagnetic simulations.

**Author contributions**

J.G. designed the experiments and supervised the study. A.L. and M.M. grew the LT-GaAs substrates and optimized their properties. M.H. performed the magnetic sample deposition and optimized the magnetic properties. E.D. fabricated the devices with help from A.A. and H.S. E.D. performed the ultrafast SOT experiments, the characterization of the ps electrical pulses, and the SOT experiments for longer pulses. J.G. performed the macrospin simulations. P.O-R. and J.G. performed the micromagnetic simulations with equipment provided by P.P. A.A. performed the static transport measurements with help from H.D., and analyzed the data to extract the simulation parameters. E.D. analyzed the SOT experimental data with input from J.G. E.D. and J.G. wrote the manuscript with input from all authors.

## Competing Interests Statement
The authors declare no competing financial interests.

## Figure captions

**Fig. 1 | Experimental setup scheme and measured pulses (ps & ns). a.** Configuration for injection and measurement of ps-wide pulses into the coplanar waveguide and transmission to the magnetic sample. On the left, a high frequency probe (copper colored) is used to make electrical contact with the sample's waveguide and provide a bias voltage ($V_b$). On the right, the red section of the waveguide depicts the position of the magnetic sample. The red shades represent the laser beams irradiating the photoswitches. The generated ps-wide electrical pulse is represented by a green trace. The lock-in current detection is represented by an amperemeter (A). Inset, from left to right: emitter photoswitch (scale bar: 50 μm), detector photoswitch (scale bar: 20 μm), and magnetic sample (scale bar: 5 μm). **b.** On-chip measurement of various pulses, all of them generated with a 5 V bias voltage. The resulting pulse duration can be tuned by changing the laser pump fluence, as seen in Extended Data Fig. 2. **c.** Configuration for injection and measurement of pulses >0.5 ns from a commercial pulse generator via the high frequency electrical probe. The generated ns-wide electrical pulse is represented by a purple trace. **d.** On-chip measurement of a 5-ns pulse with a nominal amplitude of 2.7 V.

**Fig. 2 | Switching dependence on current density and in-plane field strength. a.** Switched area as a function of normalized current in the *Co* sample, for SOT induced by 7-ps and 1-ns pulses. The averaged switched area was defined as the area percentage that is switched after one pulse, averaged over several events (Suppl. Mat. S3). The current was normalized by its value at the threshold (100 % switched). In-plane field intensity was 160 mT for all points. The error bars correspond to the standard deviation of the switched area percentage. **b.** Normalized threshold current as a function of in-plane field strength in the *Co* sample, for SOT induced by 7-ps and 1-ns pulses. Current was normalized by the current threshold at 160 mT. The error bars are standard deviations derived from the error in pulse calibration.

**Fig. 3 | Switching dependence on current pulse duration. a.** Critical current as a function of pulse duration for all three samples. The colored solid lines represent the $1/\tau_P$ current dependence for pulses ≲10 ns in the *Co* and *CoGd* samples, while the light gray line indicates the $-\log^{0.5}(\tau_p)$ current dependence for longer pulses. The dashed lines are guides for the eyes obtained with the sum of a $1/\tau_P^{0.5}$ function and a Lorentzian centered around the ferromagnetic resonance period. The error bars are derived from the error in pulse calibration. The right axis shows the estimated current density flowing in the 4-nm Pt layer within the *Co* stack, calculated with a parallel resistor model (see Suppl. Mat. S7). **b.** Energy cost of switching as a function of pulse duration for all three samples. The yellow star marks the all-optical magnetization switching (AOS) of the *CoGd* sample, achieved with a ~150 fs optical pulse. The dashed and solid lines correspond to the $1/\tau_P$ dependence and the guide for the eyes from panel **a**, respectively. The error bars associated to pulse duration are contained within the symbols and are thus not shown. The error bars are standard deviations derived from the error in pulse calibration. The right axis shows the energy cost projection to a 100 x 100 nm² device.

**Fig. 4 | Comparison of micromagnetic and macrospin simulations in the ns and ps regimes. a.** Micromagnetic simulation of the evolution of the reduced out-of-plane component of the magnetization in the *Co* sample induced by a 2-ns pulse with $j = 1.2 \times 10^{12}$ A m⁻², arriving at $t_0$ = 2.5 ns. **b.** Micromagnetic simulation of the evolution of the reduced out-of-plane component of the magnetization in the *Co* sample induced by a 6-ps pulse with $j = 8.5 \times 10^{12}$ A m⁻², arriving at $t_0$ = 20 ps. **c.** Micromagnetic and macrospin-simulated dynamics of the spatially-averaged out-of-plane magnetization ($m_z$) induced by a 2-ns pulse of $j$ =1.2 x 10¹² A m⁻² and $j$ = 1.5 x 10¹² A m⁻², respectively. The 2-ns pulse profile is depicted below in arbitrary units. **d.** Micromagnetic and macrospin-simulated dynamics of $m_z$ induced by the 6-ps pulse of $j$ = 8.5 x 10¹² A m⁻² and $j$ = 8.6 x 10¹² A m⁻², respectively. The 6-ps pulse profile is depicted below in arbitrary units. **e.** Macrospin dynamics of the spatially-averaged reduced in-plane magnetization components ($m_x, m_y$) induced by the 2-ns pulse. **f.** Micromagnetic dynamics of $m_x$ and $m_y$ induced by the 2-ns pulse. **g.** Macrospin dynamics of $m_x$ and $m_y$ induced by the 6-ps pulse. **h.** Micromagnetic dynamics of $m_x$ and $m_y$ induced by the 6-ps pulse. All simulations were performed under an in-plane field of 150 mT. For further details see Suppl. Mat. S10.

**Fig. 5 | Comparison of experimental and simulated critical current densities and energies as a function of pulse duration.** The micromagnetic simulation follows a $1/\tau_p^{0.5}$ law that correctly reproduces the current densities and energy consumption for 7-ps and 1-ns pulses, but do not fit the full datasets. The $1/\tau_p$ law is also shown for reference. Macrospin simulations are shown, both without temperature dependence (no T) and including a temperature dependent anisotropy (thermal). The experimental error bars, derived from the error in pulse calibration, are contained within the symbols and are thus not shown.

## Methods

**Sample fabrication.** The devices were fabricated over a substrate of LT-GaAs grown by molecular beam epitaxy (MBE) in three steps. First, we grew a 400 nm layer of high-temperature GaAs over a GaAs (100) commercial substrate, followed by a 300 nm layer of $Al_{0.33}GaAs_{0.67}$, which acts as an insulator between the high-temperature GaAs and the next step. In the second step, we deposited the photoactive layer of the substrate, that is, a 1-µm layer of LT-GaAs at 260°C.

In the third and final step, the substrate was annealed ex-situ at 575°C in an ArH$_2$ atmosphere in order to increase its resistivity by means of reducing the number of conduction states in the semiconducting gap due to defects.

Device fabrication was done through a four-step UV lithography recipe. First, a deposition of Ti(10 nm)/Au(20 nm)/Ti(3 nm) was made and patterned into the detector structure employed to measure and calibrate the electrical pulses. Second, a layer of 100 nm of SiO$_2$ is deposited over the entirety of the sample in order to electrically insulate the detector structure from the CPW, and also to avoid current leakage through the semiconducting substrate between the signal and ground lines of the CPW. Holes were patterned into the SiO$_2$ layer in order to define both generator and detector photoswitch, and also around the contact pads of the detector structure in order to establish proper ohmic contact with the GSG tips. Third, the magnetic stacks were grown by dc-magnetron sputtering in an AJA system, and patterned into 18 x 4 μm$^2$ strips distributed through the substrate. Finally, in the fourth lithography step, we deposited a bilayer of Ti(10 nm)/Au(300 nm) and patterned the CPW, which includes the embedded photoconductive switch and a tapered zone in which a 3-μm gap in the transmission line is aligned with the magnetic strip of the previous step, leaving a 3 x 4 μm$^2$ effective magnet into which current is injected.

The magnetic samples were embedded at the end of the waveguides as impedance-matched resistive loads. Impedance matching yields strong pulse energy absorption, therefore high efficiency (see Extended Data Fig. 3For that reason, due to the stack resistivity, short (~3 μm) and wide (~4 μm) devices were fabricated.

**Optical setup.** The magneto-optical experiments were performed using the amplified ytterbium laser system PHAROS® (Light Conversion) and the optical parameter amplifier ORPHEUS-F® (Light Conversion), which can produce ~200 fs pulses centered around 785 nm. The laser output goes through a beamsplitter and is divided into two beams: an s-polarized beam constituting the pump, and a p-polarized probe. The pump goes through a 13-ns delay line and is subsequently focused on the generator photoswitch in an 80-μm wide spot (FWHM). For ps-pulse detection and characterization, the probe is focused into a 25-μm spot through a 2x objective onto the detector. A repetition rate between 1-5 kHz was used to avoid static heating and the breakdown of the waveguides due to intense electric fields. For single-pulse switching experiments the laser was operated in single-shot mode, and the probe beam was kept blocked.

The sample magnetic state was monitored using a home-built magneto-optical Kerr effect (MOKE) microscope using a 628-nm LED light source and a 50x objective (numerical aperture of 0.42), yielding a resolution of about 1 μm.

**SOT switching procedure.** The switching threshold current was determined as follows. The photoswitch was set at a bias voltage ($V_b$) and an in-plane field was applied to the sample. We checked that the in-plane field did not affect the magnet's final state (due to parasitic out-of-plane components for example). Therefore the field only helps slightly tilting the magnetization away from the easy axis, as is required for SOT switching[27]. A single ps-wide electric pulse was generated by exciting the photoswitch with a laser pulse, which was then transmitted through the waveguide and injected into the magnetic sample. This pulse induces SOT and also heats up the sample, which reduces its anisotropy field and therefore assists magnetization reversal[15]. The magnetic state was monitored in-situ by MOKE microscopy. $V_b$ was increased until a pulse-induced switching event was observed. Then, additional pulses were generated in order to confirm the process deterministic nature. As a final test, SOT was confirmed by repeating the experiment with reversed current pulse polarity (i.e. electrical bias) and in-plane field orientation. The same procedure was followed for 1-ns pulses as well, in which case the current was swept by changing the pulse generator output amplitude.

**Critical current and energy definitions.** The current is given by $I = V(1 - \Gamma)/Z_0$, where $V$ is the calibrated voltage measured by our THz detection system, $\Gamma$ is the reflection coefficient (determined as indicated in Suppl. Mat. S1, see Extended Data Fig. 3) and $Z_0 = 50\ \Omega$ is the waveguide impedance. The energy cost was calculated as the integral over time of the delivered power, given by $P(t) = V(t)^2(1 - \Gamma^2)/Z_0$, where $V(t)$ is the time-resolved trace of the incident pulse.

**Model dependance on parameters and pulse shape.** We note that our macrospin simulations are highly dependent on the various parameters and their temperature dependencies (magnetization, exchange, Dzyaloshinskii-Moriya interaction (DMI), anisotropy, and so on). Additionally, we believe the shape of the pulse might be quite important. Experimentally, as shown in Extended Data Fig. 2 and Extended Data Fig. 4, pulses from different regimes of pulse duration have different shapes: Gaussian, asymmetric Gaussian, quasi-square, and square. In our simulation work, however, we have assumed only Gaussian pulses. The shape of the pulse might change heating, nucleation probability and critical currents, and should be accounted for in future works. However, as a general observation, a coherent reversal was always obtained in the ps case, whereas for the ns case both almost coherent reversal and nucleation-dominated reversal are possible. For all pulse durations, if the current is strong enough, magnetization can be forced coherently across the equator, although at the expense of energy efficiency.

### Data Availability Statement
Data is available from the corresponding authors upon request.

### Code Availability Statement
Codes are available from the corresponding authors upon request.

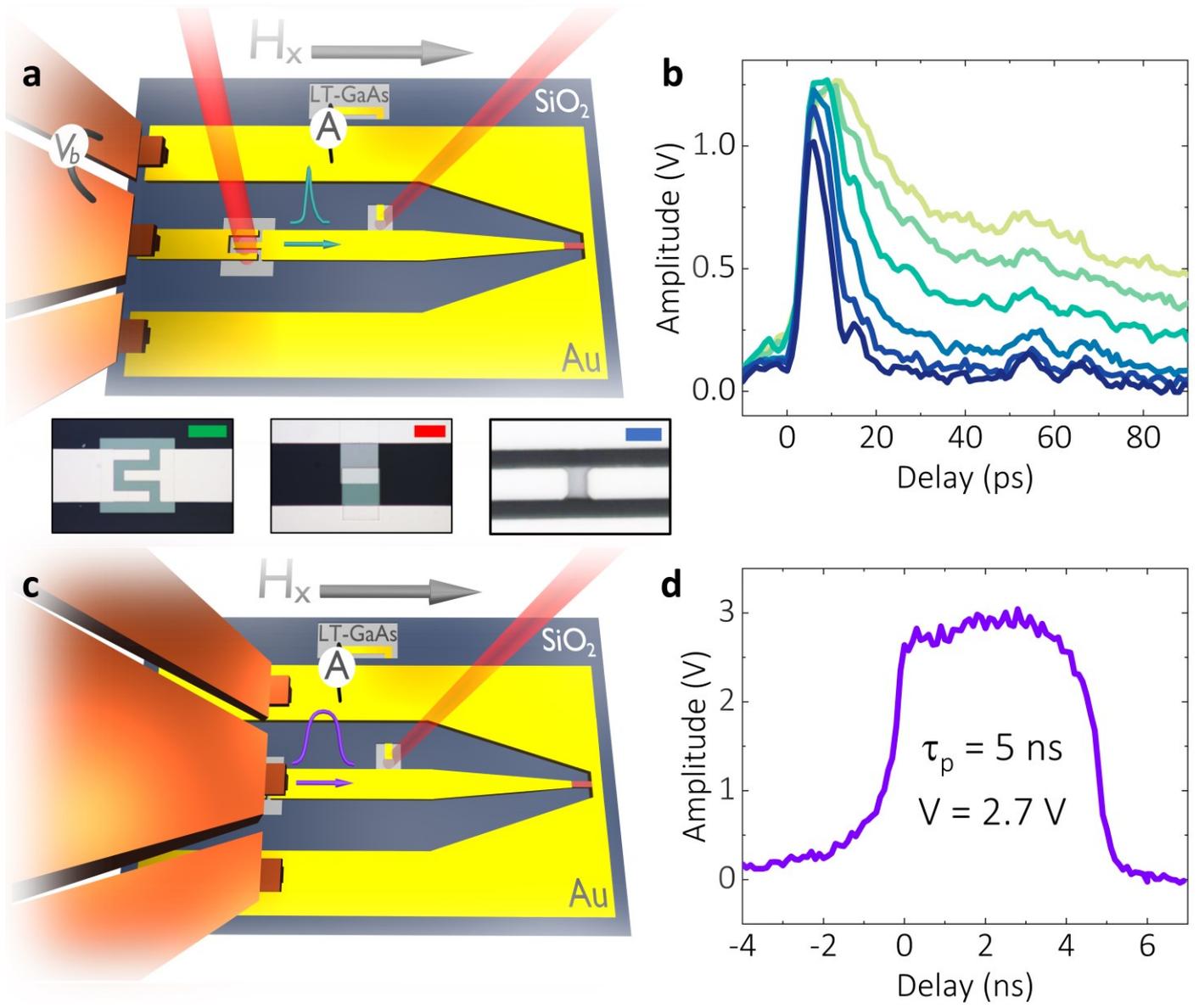

Figure 1

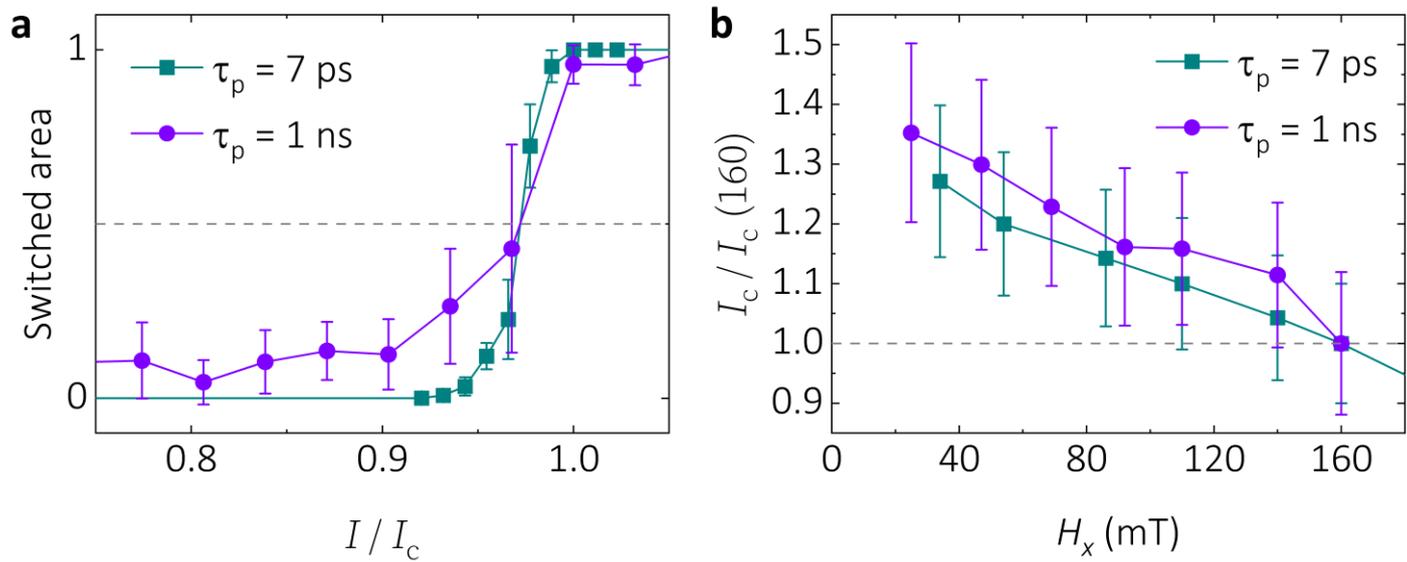

Figure 2

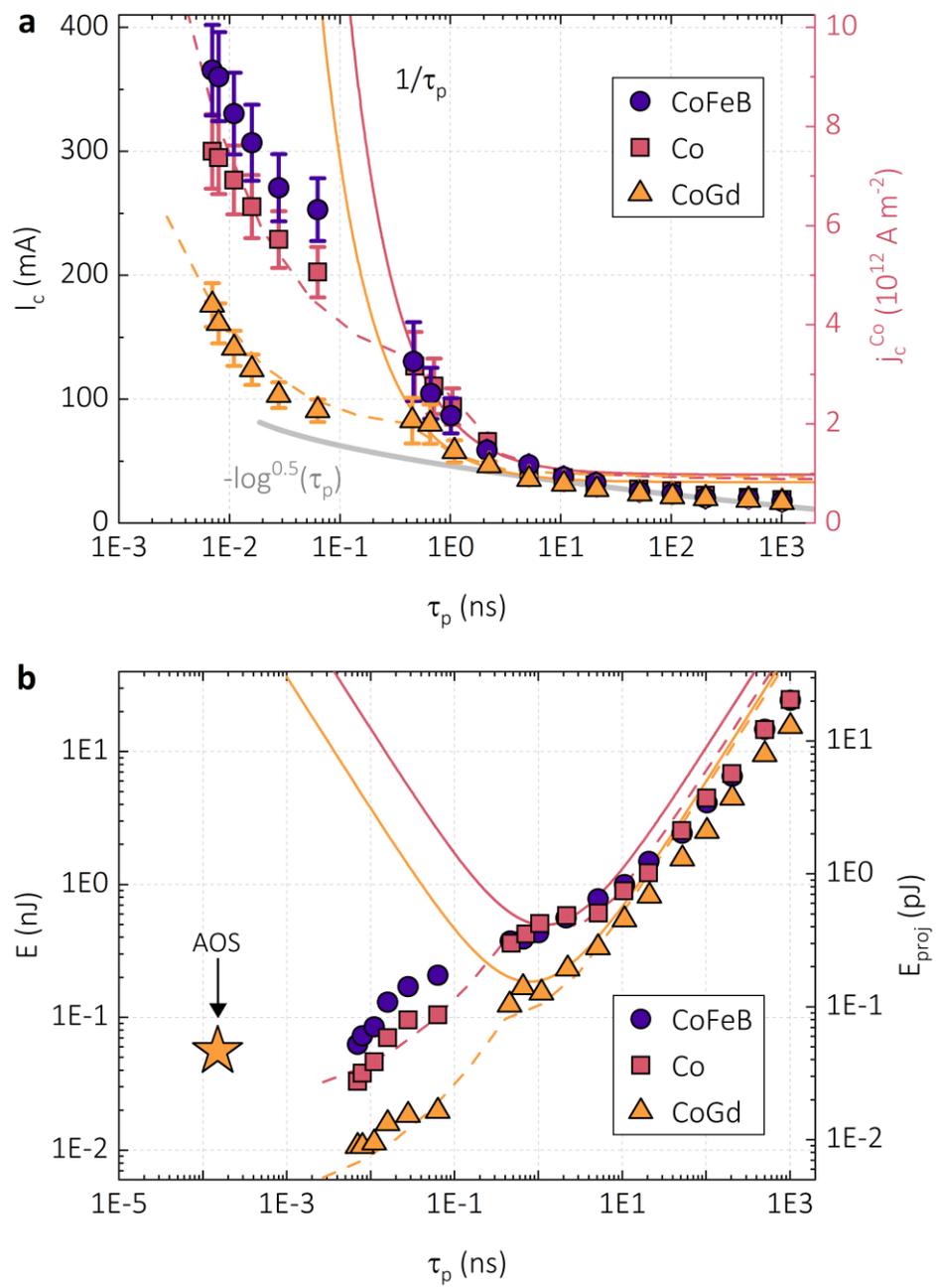

Figure 3

**ns dynamics**              **ps dynamics**

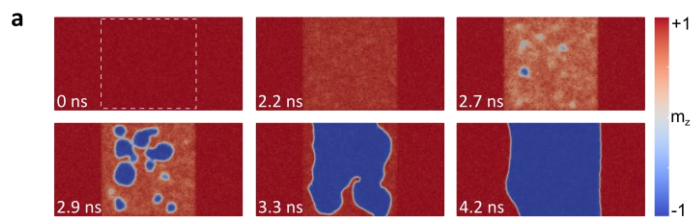 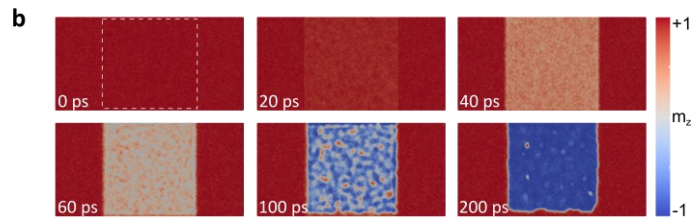

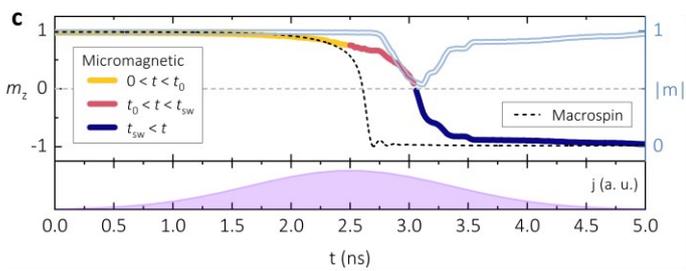 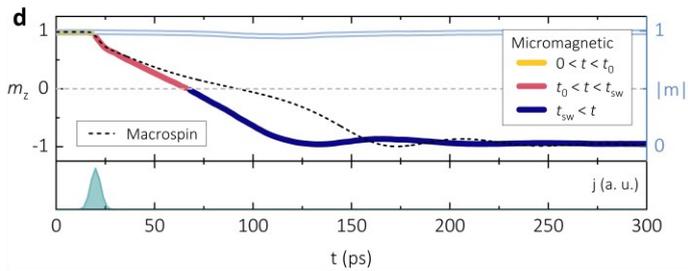

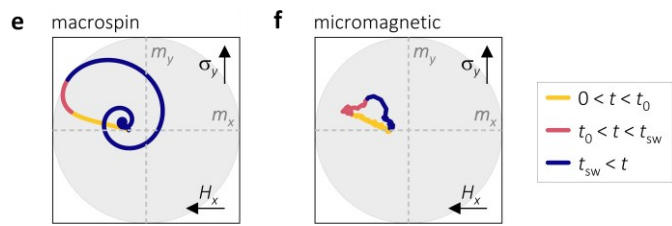 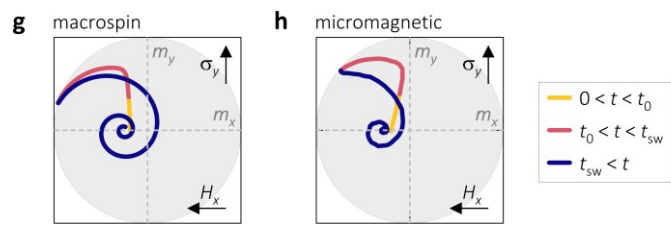

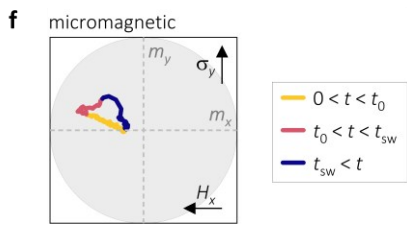 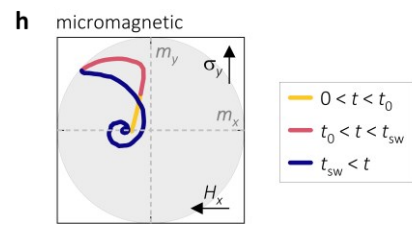

Figure 4

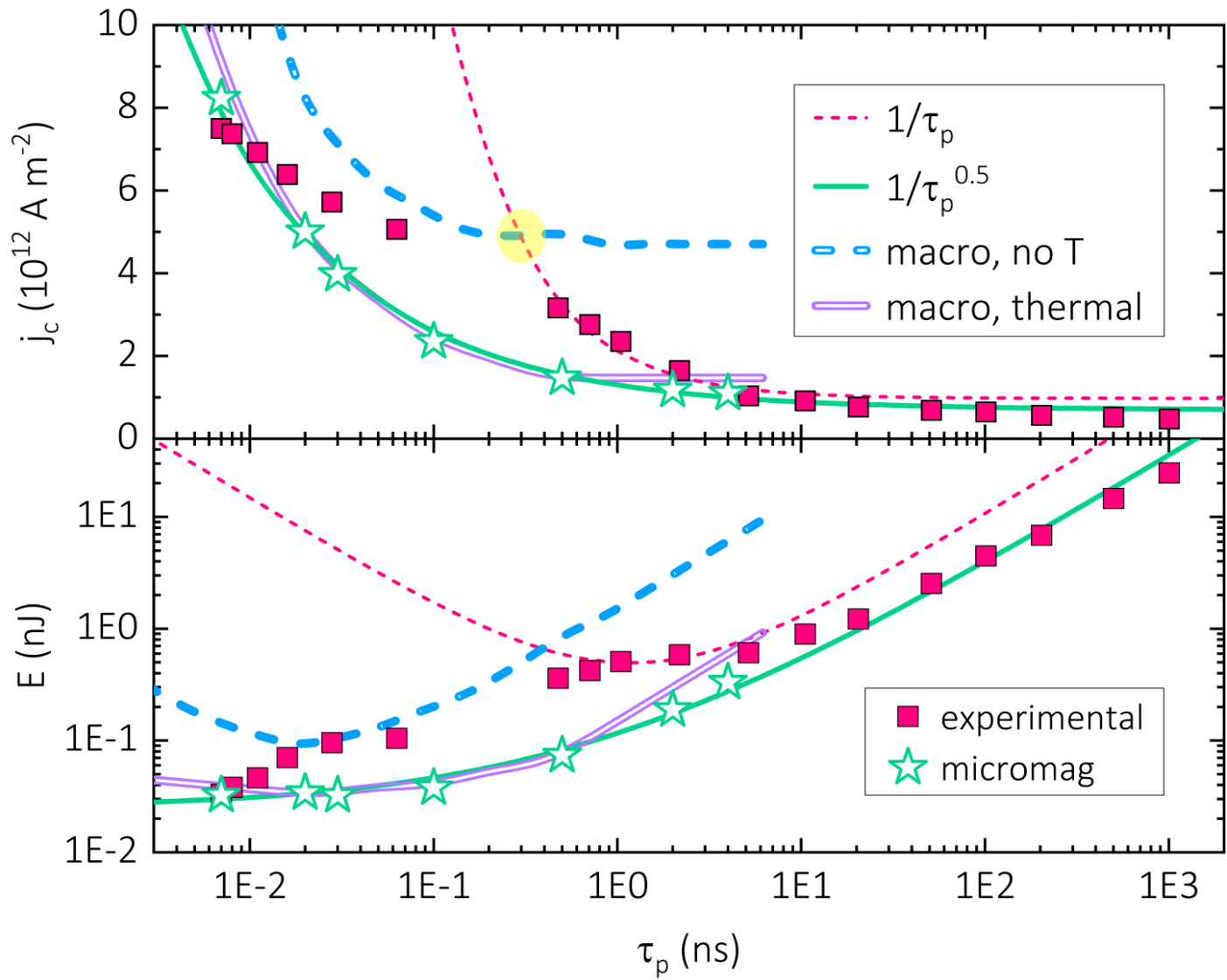

Figure 5